\documentclass[preprint]{aastex}

\shorttitle{Damp Mergers and Globular Clusters}
\shortauthors{Forbes, Proctor, Strader \& Brodie}

\def\etal{{\it et al.~}}

\begin{document}

\title
{Damp Mergers: Recent Gaseous Mergers without Significant 
Globular Cluster Formation?}
\author
{Duncan A. Forbes\thanks{dforbes@astro.swin.edu.au}, Robert Proctor, 
Jay Strader, Jean P. Brodie\\
Centre for Astrophysics \& Supercomputing, Swinburne University,
  Hawthorn, VIC 3122, Australia\\
UCO/Lick Observatory, University of California,
 Santa Cruz, CA 95064, USA}


\begin{abstract}

Here we test the idea that new globular clusters (GCs) are formed
in the same gaseous (``wet") mergers or interactions that give rise
to the young stellar populations seen in the central regions of
many early-type galaxies. We compare mean GC colors with 
the age of the central galaxy starburst. 
The red GC subpopulation reveals remarkably constant mean
colors independent of galaxy age. 
A scenario in which the red GC subpopulation is a
combination of old and new GCs (formed in the
same event as the central galaxy starburst)  
can not be ruled out; although this would require an age-metallicity
relation for the newly formed GCs that is steeper than 
the Galactic relation.
However, the data are also well described 
by a scenario in which most red GCs are old, and few, if any, are
formed in recent gaseous mergers. This is consistent with the
old ages inferred from some spectroscopic studies of
GCs in external systems. The event that induced the central
galaxy starburst may have therefore involved insufficient gas
mass for significant GC formation. We term such gas-poor
events ``damp" mergers.

\end{abstract}


\section{Introduction}

Since the first hints that the color distributions of globular
cluster (GC) systems in some early-type galaxies were bimodal
(Couture, Harris \& Allwright 1991; Ashman \& Zepf 1993), 
bimodality has been shown to be the norm (e.g., Larsen \etal 2001; 
Kundu \& Whitmore 2001; Strader \etal 2006; Peng \etal 2006). 
This blue/red bimodality indicates distinct
metallicity (and perhaps age) subpopulations within the GC
system. The mean color of the blue and red subpopulations are known 
to correlate with their host galaxy luminosity (Strader, Brodie \& 
Forbes 2004). 
Basic interpretations of these two subpopulations include dissipative 
formation
at early epochs (Forbes, Brodie \& Grillmair 1997), gaseous mergers at
early or late epochs (Ashman \& Zepf 1992), 
and the dissipationless accretion of GCs (Cote, West \& Marzke
2002). A key piece of evidence in assessing these scenarios is
the age distribution of the two GC subpopulations.
The current state of affairs is summarised
in the review of Brodie \& Strader (2006), which notes that small spectroscopic
samples to date find most GCs to be $\ga 10$ Gyrs old, implying an early 
formation
epoch at redshifts $z \ga 2$.
However, such work is limited by sampling biases: only the brightest
GCs  
in nearby ($\la 25$ Mpc) galaxies can be studied.  
To create large samples of \emph{typical} GCs,
one must use photometric studies.

Many early-type galaxies are found to have younger central stellar
populations, indicative of a recent interaction or gaseous (``wet")
merger which induced some star formation (Trager \etal 2000;
Proctor \& Samson 2002; Terlevich \& Forbes 2002;
Thomas \etal 2005; Denicolo \etal 2005; Sanchez-Blazquez \etal
2006, SB06).\footnote{In some low-mass galaxies, ``cold" flows
from the intergalactic 
medium may be a possible source of gas (e.g., Keres \etal~2005).} The
age of this young starburst has been found to correlate with a
galaxy's location on the Fundamental Plane and with morphological 
fine structure (Schweizer \& Seitzer 1992; Forbes, Ponman \& Brown
1998).
There is also  good evidence for the
formation of proto-GCs in ongoing wet mergers like the Antennae
(Whitmore \& Schweizer 1995), which are thought to ultimately form an
elliptical galaxy (e.g., Toomre \& Toomre 1972). 
Furthermore, as star clusters appear to be the
primary mode of star formation (Lada \& Lada 2003), we expect GCs
to trace major star formation episodes in all starbursts (Larsen
\& Richtler 2000). In the semi-analytic models of Beasley \etal
(2002) metal-poor GCs form in proto-galactic fragments at high
redshift, while metal-rich ones form in subsequent gas-rich merging events. 
However, if formation
conditions are not suitable (e.g., insufficient gas mass or 
low GC formation efficiency) then few long-lasting GCs may be produced. 
Local examples of this include the Galaxy's disk and the
LMC, which have experienced open cluster formation without GC formation. 
Also, in the case of gas-free (``dry'') accretions or mergers no   
induced star, or star cluster, formation is expected. 

Major gaseous mergers (Ashman \& Zepf 1992) are predicted 
to form new metal-rich GCs with a mean age corresponding to that of the 
merger. Any new GCs formed 
will add to the existing GC system of the progenitor 
galaxies, i.e. an old metal-poor (blue) and possibly an old but more 
enriched (red) subpopulation. 

The newly formed metal-rich GC subpopulation will initially be
very blue, however it will redden rapidly (e.g., Whitmore \etal
1997). If formed in the last Gyr, the subpopulation may have a
color similar to that observed for blue subpopulations in
early-type galaxies. For the first couple of Gyrs, depending on
its metallicity, it may have intermediate colors. However, older 
than a few Gyrs, the colors resemble
those of red subpopulations, and continue to redden slowly with
age. If significant numbers of new GCs were formed more than a
few Gyrs ago, then the red subpopulation of the merger remnant galaxy
may be dominated by these newly formed metal-rich GCs.  The mean
color of the red subpopulation will therefore be determined by
the relative fraction of newly formed vs old  GCs and
their respective enrichment levels (metallicity) and time elapsed
(age) since formation.

Although broad-band colors are subject to the age-metallicity
degenearcy, the evolution of color for any reasonable GC
metallicity has a similar characteristic form and can be well
described by a single stellar population model. Thus we can model
the expected mean color of the red subpopulation with age and
compare this to the observed color and age of the galaxy's
central starburst. Here we examine two relatively large samples
of early-type galaxies in order to test the idea that the red
subpopulation of GCs were largely
formed in gaseous mergers (Ashman \& Zepf
1992).


\section{The Data}

We present two datasets for which we have obtained GC colors
and central starburst ages for the same galaxies. The first 
sample has been gathered from a wide variety of 
literature sources; we refer to this as our heterogeneous sample. 
The second sample comes from a single photometric
source (the Virgo ACS Survey; Peng \etal 2006) and single galaxy age
study (Caldwell \etal 2003); we refer to this as our
homogeneous sample.

\subsection{A Hetrogeneous Sample: Galaxy Ages}


Stellar age estimates based on 
an analysis of Lick absorption
lines are available for a number of early-type galaxies. Such ages
are luminosity weighted, and date the last major starburst
while giving little information on the mass involved. They are
also central values which may not therefore reflect the galaxy as a whole 
(see, e.g., discussion in Terlevich \& Forbes 2002). 

For our first sample of 36 galaxies, we take galaxy ages from four
independent studies: Terlevich \& Forbes (2002), Thomas 
\etal (2005), Denicolo \etal (2005), and SB06. All
four employ a similar method of using the H$\beta$ index and the
[MgFe] composite index, together with a single stellar
population (SSP) model, to derive galaxy ages
and metallicities. In Table 1 we list the central ages from 
these four studies for those galaxies which also
have GC colors available from the
literature. When more than one age estimate is available, the
agreement among studies is generally quite acceptable.
For these galaxies we adopt the average age value. 
However, for four galaxies the estimates are wildly different.

The galaxies with discrepant age estimates are:\\ 
NGC 1052: We adopt the young age of Denicolo \etal (2005). 
This is supported by Pierce \etal (2005), who measured a
galaxy age of $\sim 2$ Gyrs.
NGC 1407: 
We adopt the old age given by Thomas \etal (2005). This is
supported by the high-quality spectra of 
Cenarro \etal (2006) and Spolaor \etal (2006) which 
indicate very old central ages. 
NGC 3115: 
We adopt the age of SB06, but note that 
Norris \etal (2006) have shown
that NGC 3115 has a younger age along the major than minor axis.
NGC 4365: 
We adopt the 7.9 Gyr age of SB06 which is
supported by the 9.7 Gyr age derived by Proctor \& Samson (2002).
Our final adopted age is given in the last column of Table 1.

Using all the galaxies with multiple age estimates, we measure an
average rms scatter of $\pm$0.2 dex in log age (ages are derived
from SSP model grids which are nearly uniform in log), or $\sim$2 Gyrs.
A further indication that our age estimates are reasonable is
provided by Fig. 1. Here we show the residual from the
Fundamental Plane value for each galaxy (from Prugniel \& Simien
1996) in log units against log of our adopted galaxy age for the
available galaxies. The solid line is the fit from a much larger
sample of Forbes, Ponman \& Brown (1998). The data scatter about
the Forbes \etal fit line showing that the galaxy ages are
related to an independently measured value, i.e. the deviation of
the galaxy from the standard Fundamental Plane. The rms scatter
in galaxy age is consistent with the 
value quoted above for age estimates between different
studies.  Finally, we note
that excluding the four galaxies with the most discrepant ages in
Table 1 from the analysis 
does not affect our results. 

\subsection{A Heterogeneous Sample: Globular Cluster $V-I$ Colors}

A large number of early-type galaxies have been imaged with sufficient
precision to detect bimodality in the GC color distribution,
particularly with the {\it Hubble Space Telescope}.  Although ACS imaging of
GCs is becoming more common (e.g., Peng \etal~2006; 
Harris \etal 2006; Forbes \etal
2006; Strader \etal 2006; Spitler \etal 2006), most HST
studies have used the WFPC2
camera. These latter data were typically taken in the V and I
filters. We have searched the literature to compile a database of
mean blue and red GC colors for early-type galaxies observed by HST or with
ground-based telescopes. It excludes the recent Virgo ACS survey
which is discussed in section 2.3. We do however include the
Sombrero Sa galaxy as it is extremely bulge dominated.  
We have transformed other filter combinations (i.e. mostly B--I
and C--T1) 
into V--I using the conversions of Forbes \& Forte (2001). 
The conversion of Forbes \& Forte (2001) is an empirical one based on
the Galactic GC system, i.e. it assumes old stellar populations. 
A conversion for younger stellar populations would predict a
bluer V--I color than an assumed old age. However 
the differences are very small because the filters are similar to
V--I and the age-metallicity degeneracy means tracks of different
ages run roughly parallel to each other. This is illustrated in
Fig. 2 which shows Bruzual \& Charlot (2003; BC03) SSP model tracks for single ages of 2 and
12.5 Gyrs at various metallicities. Over the typical B--I color
range of GCs (i.e. 1.5 $<$ B--I $<$ 2.2) the predicted V--I
color is $\le$ 0.03 mag. redder if GCs are as young as 2 Gyrs
compared to our assumption of 12.5 Gyrs. A similar offset occurs
if C--T1 is compared to V--I. 
As well as these filter transformations, 
we have applied extinction corrections from Schlegel \etal (1998) to
all GC colors. In Table 2 for each galaxy with an age estimate, 
we list the 
(V--I)$_o$ colors after transformation from the original filter
combination and extinction correction. We also give the original
reference for the GC photometry, whether the bimodality is
confirmed (Y) or likely (L), the original filter combination and
whether the photometry comes from HST (Y) or not (N). A complete
table of GC photometry (i.e. including those galaxies without age
estimates) can be found at:\\ 
http://astronomy.swin.edu.au/dforbes/colors.html\\

The galaxies in this sample cover a range of luminosities, and 
GC colors are known to correlate with the luminosity of their host
galaxy (see Brodie \& Strader 2006 for a review). This correlation is thought 
to be largely a metallicity-mass relation. 
Before searching for any correlations of GC color with galaxy age,
we have removed this effect using the observed GC V--I color--galaxy
luminosity relations of Strader, Brodie \& Forbes (2004). 
This correction (of typically less than 0.05 mag.) 
eliminates, at least to first order, 
any GC color variations due the range of
host galaxy luminosities in our sample. 

\subsection{A Homogeneous Sample: Galaxy Ages}

Central ages for 33 low-luminosity early-type 
Virgo galaxies are taken from a single source, i.e. Caldwell 
\etal (2003). We use their best fit age, which comes 
from comparing high order
Balmer lines and Fe4383 with an 
SSP model. Using a single source avoids any systematic
differences that may be present between multiple sources.

\subsection{A Homogeneous Sample: Globular Cluster g--z Colors}

Mean colors for the blue and red GC subpopulations of early-type
Virgo galaxies come from the ACS Virgo Survey of 
Peng \etal (2006). This dataset has several advantages over the
V--I colors from the literature; namely g--z has twice the
metallicity sensitivity of V--I, the photometric uncertainties
are generally lower and galaxies are at the same distance so
any potential aperture effects do not play a role.  
Uncertainties in the peak colors range from $\pm$0.01 to
0.3 mag.  Following Peng \etal we assume m--M = 31.09 to the Virgo cluster. 

For VCC 1488 (IC3487), the mean blue color 
listed in Peng \etal (2006) of g--z = 0.72 does not seem to correspond to the
fitted Gaussians displayed in their Figure 2. In this case we have
chosen to take the mean color for the blue GCs from the work of
Strader \etal (2006). This value, of g--z = 0.82, better represents the
color distribution. 
Finally, we correct all GC g--z colors
using the GC color--galaxy luminosity relation of Peng \etal (2006).

\section{Results and Discussion}

Having described the data from which we take GC mean subpopulation
colors and central galaxy ages, we now present our results. As well
as the color of the red subpopulation we show results for the color
difference between the red and blue subpopulations. A variety of
observational and theoretical work suggests that blue GCs may
have been formed
before reionization at z $\ge$ 6. This implies ages of $\ge$12.7
Gyrs.  Such old ages are reflected in the observed near constant
color of blue GC subpopulations. The color difference is therefore
largely due to the age and metallicity of the red GCs, and 
it may be more accurately determined than the
individual subpopulation mean colors.

Figure 3 shows the mean V--I and g--z colors of the red GC
subpopulations, and the color difference between red and blue
subpopulations versus galaxy stellar age for our two 
samples. The GC colors in this, and subsequent figures, have been
corrected for the GC color--galaxy luminosity relation as
described above. 

For the hetreogeneous sample we find constant
V--I colours with an rms spread of 0.044 mag., or an error on the
mean for 36 galaxies of 0.007 mag. For the homogeneous sample we find constant
g--z colours with an rms spread of 0.068 mag., or an error on the
mean for 33 galaxies of 0.012 mag. 
The rms scatter is equal to, or less than, the typical
photometric errors in our samples suggesting that the intrinsic scatter is
indeed very low.

In Fig. 3 we use different symbols within the heterogenous sample
when the bimodality is classified as confirmed (Y) or likely (L) in Table
2. The figure shows that there is no systematic offset between
the two classifications of bimodality; for subsequent figures we
use a single symbol for the heterogenous sample. 
An analysis of a subsample of the ACS Virgo Survey was first
carried out by Strader \etal (2006). A comparison of their
sample with that from Caldwell \etal (2003) reveals 12
galaxies in common with GC color and galaxy age measurements. We have
examined them in a similar way to above, finding constant g--z
colors, i.e. an rms scatter of 0.059 and an error on the mean of 0.017 mag.

In this and subsequent figures we show 
evolutionary tracks based on SSP models 
from Bruzual \& Charlot (2003) which assume a
Salpeter initial mass function and a solar abundance ratios  
(we note that other SSP models
give similar results).
  
The track in Figure 3 is the scenario in which {\it all} red GCs
are uniformly old with a fixed metallicity of [Fe/H] = --0.7 and
hence the evolutionary track has a constant color. This is the
scenario in which no GCs formed at the time of the central
starburst so that GC
color is unrelated to galaxy age.  We call this the no evolution
(NE) track and it provides a good fit to the mean color of the
red subpopulation. For the color difference between the red and
blue subpopulations we have simply shifted the NE track bluer by
an arbitrary amount. Again it provides a good representation of the
data.

So although the scenario in which all of the red GCs are old, and none are 
formed in recent mergers, is a good fit to the observed data can 
other scenarios be ruled out ?

Figure 4 shows 
the predicted change in color for a fixed metallicity
(FM) of [Fe/H] = --0.7. This track corresponds to the situation
in which {\it all} red GCs in a galaxy formed at the time of the
central galaxy starburst with a metallicity of [Fe/H] = --0.7
(i.e. the metallicity required for very old GCs to be consistent
with their measured colors). It provides a plausible fit to the
data at old ages but deviates strongly from the data for young
ages. Lower metallicity tracks would provide a worse fit at all
ages. Higher metallicity (e.g. solar) tracks could explain the GC
colors in the youngest galaxies but would fail to match the
color of the oldest galaxies. Thus we can fairly confidently
rule out the scenario in which all red GCs were formed in a merger
from gas of a fixed metallicity over a wide range of epochs. 

A more plausible scenario is that the gas from which GCs form is
steadily enriched over time, so the oldest red GCs may have formed from
[Fe/H] = --0.7 gas (to match the observed color) but younger
GCs, forming more recently, would be
more metal-rich. Thus we need to assume a relationship for 
the enrichment of the gas with time (age) as the Universe
evolves.
A plausible age-metallicity relation comes from the fossil record
of the Galactic disk. From studies of individual stars and star
clusters, Edvardsson \etal (2003) find that a simple linear relation between 
metallicity and log age is a reasonable fit to the data (albeit with some
scatter). Thus for this scenario we assume 
{\it all}
red GCs formed at the time of the central galaxy starburst with
a metallicity determined by the Galactic age-metallicity relation.
Figure 5 shows this mixed
metallicity (MM) evolutionary track. 
This track provides a good fit to the majority of the data.
A fit
of the data points to the MM track gives an rms of 0.056 for the
hetreogenous sample and 0.077 for the homogenous sample, 
i.e. similar to that of the NE track. A slightly steeper
age-metallicity relation (producing redder GCs at a given age) 
would provide a better fit; conversely
a shallower age-metallicity would provide a worse fit to the data. 

The mixed metallicity track assumes 100\% of the red GCs are
formed at the time of the recent central starburst. A contribution of
old, metal-rich red GCs from the progenitor galaxies will tend to
`dilute' the track making it more similar to the no evolution
track of Figure 3. 
Perhaps a good illustration of this is the young merger remnant NGC
3610 in our heterogeneous sample. 
This galaxy has a central starburst age of $\sim1.8$
Gyr.  Strader, Brodie \&
Forbes (2004) obtained spectra of 10 red GCs and found only 2
to have an age consistent with forming at the same time as the
galaxy starburst. 
If these 10 GCs are representative then
80\% of the red GC subpopulation is old and 20\% formed in a
gaseous merger event $\sim$ 1.8 Gyrs ago.  
The red GC subpopulation has a mean V--I color that is
consistent with the pure no evolution track but also with one
that contains a 20\% contribution from our mixed metallicity
track. Thus a contribution of old and newly formed red GCs can
account for the observed GC colors in the youngest
galaxies if the age-metallicity relation is not shallower than 
assumed. 

\section{Concluding Remarks}

We have compared the observed mean color of the red GC
subpopulation and its color difference with the blue
subpopulation to the age of the central starburst for two samples
of early-type galaxies. We find that, for both samples, the GC
subpopulation colours show no trend with galaxy age and
effectively have a constant value. To interprete this result,  
three evolutionary scenarios are discussed. 
A fixed
metallicity for newly formed red GCs is a poor representation of the data,
especially for young galaxy ages. 
A mixed metallicity scenario (in which the younger GCs form from
progressively 
more metal enriched gas)  
provides a better fit to the data. 
The fit can be improved with a steeper
age-metallicity relation than we have assumed and/or a
significant contribution from old GCs to the red GC subpopulation in
young galaxies; a shallower age-metallicity relation 
is inconsistent with the data.
A good fit to the data is provided by 
a no evolution scenario in which most of the red GC are 
uniformly old, and few, if any, formed 
in late epoch gaseous mergers.

This latter conclusion 
supports some previous spectroscopic work that most GCs in
elliptical galaxies are old, and that few new GCs survive from 
the same gaseous merger event that
gave rise to the central young starburst.  
There are
several possible interpretations: (i) proto-GCs
formed but were quickly destroyed, (ii) GCs formed
with unexpectedly low efficiency in major star-forming events, or
(iii) the starburst involved little gas mass,
and so few new stars and GCs were formed. We term the latter gas-poor 
events ``damp" mergers. 
If situations (i) and (ii) can be ruled out by studies of nearby ongoing
mergers, 
then the our results would be
more consistent with frosting models (e.g., Trager \etal~2000)
than with quenching models (Faber \etal~2005) for early-type
galaxies.  \\

\noindent{\bf Acknowledgements}\\
We acknowledge Francois Schweizer for making the initial
suggestion that motivated this work.  
We thank the referees for several useful comments that have
improved this paper.  
This material is based upon work supported
by the Australian Research Council and the 
National Science Foundation under Grant AST-0507729.\\

\newpage

\noindent{\bf References}

\noindent
Ashman, K., Zepf S., 1992, ApJ 384, 50 \\
Bassino, L., Richtler, T., Dirsch, B., 2006, MNRAS, 367, 156 (B06)\\
Brodie, J., Strader, J., 2006, ARAA, in press\\ 
Brown, R., Forbes, D., Kissler-Patig, M., Brodie, J., 2000, MNRAS, 317, 406 (B00)\\
Bruzual G., Charlot S. 2003, MNRAS, 344, 1000\\
Caldwell, N., Rose, J., Concannon, K., 2003, AJ, 125, 2891\\
Cenarro, J., \etal 2006, AJ, submitted.\\
Cote, P., West, M., Marzke, R., 2002, ApJ, 567, 853\\
Couture, J., Harris, W., Allwright, J., 1991, ApJ, 372, 97\\ 
Da Rocha, C., Mendes de Oliveira, C., Bolte, M., Ziegler, B.,
Puzia, T., 2002, AJ, 123, 690 (D02)\\
Denicolo, G., Terlevich, R., Terlevich, E., Forbes, D.,
Terlevich, A., 2005, MNRAS, 358, 813 (D05)\\
Dirsch, B., Schuberth, Y., Richtler, T., 2005, A\&A, 433, 43 (D05)\\
Edvardsson, B., Andersen, J., Gustafsson, B., Lambert, D.,
Nissen, P., 1993, A\&A, 275, 101\\ 
Faber S., \etal 2005, astro-ph/0506044\\ 
Forbes, D., Grillmair, C., Williger, G., Elson, R., 
Brodie, J.,  1998, MNRAS, 293, 325 (F98)\\
Forbes, D., Georgakakis, A., Brodie, J., 2001, MNRAS, 325, 1431 (F01)\\
Forbes, D., Forte, J., 2001, MNRAS, 322, 257\\
Forbes D., Brodie J., Grillmair, C., 1997, AJ 113, 1652 \\
Forbes, D., Ponman, T., Brown, R., 1998, 508, L43\\
Forbes, D., Sanchez-Blazquez, P., Phan, A., Brodie, J., Strader,
J., Spitler, L., 2006, MNRAS, 366, 1230 (F06)\\
Georgakakis, A., Forbes, D., Brodie, J., 2001, MNRAS, 324, 785 (G01)\\
Gomez, M., Richtler, T., 2003, A\&A, 415, 499 (G03)\\
Goudfrooij, P., Gilmore, D., Whitmore, B., Schweizer,
F., 2004, ApJ, 613, 121 (G04)\\
Harris, W., Whitmore, B., Karakla, D., Okon, W., Baum, W., Hanes,
D., Kavelaars, J., 2006, ApJ, 636, 90 (H06)\\
Kissler-Patig M., Richtler T., Storm M., Della Valle M., 1997, A\&A 327,
503 (K97)\\
Keres, D., Katz, N., Weinberg, D., Dave, R., 2005, MNRAS, 363, 2\\
Kundu, A., Whitmore, B., 2001, AJ, 121, 2950 (K01)\\
Lada, C., Lada, E., 2003, ARAA, 41, 57\\
Larson, S. Richtler, T., 2000, A\&A, 354, 836\\
Larson, S., 2000, priv. communication (L00)\\
Larson, S., Brodie, J., Huchra, J., Forbes, D., Grillmair, C.,
2001, AJ, 121, 2974 (L01)\\
McLaughlin, D., 1999, AJ, 117, 2398\\
Norris, M., Sharples, R., Kuntschner, H., 2006, astro-ph/0601221\\
Prugniel, P., Simien, F., 1996, A\&A, 309, 749\\
Peng, E., \etal 2006, ApJ, 639, 95\\
Pierce, M., \etal 2005, MNRAS, 358, 419\\
Proctor, R., Sansom, A., 2002, MNRAS, 333, 517\\
Sanchez-Blazquez, P., \etal 2006, astro-ph/0604568 (SB06)\\
Schlegel, D., Finkbeiner, D., Marc, D., 1998, ApJ, 500, 525\\
Spitler, L., \etal 2006, AJ, in press\\
Spolaor, M., \etal 2006, in prep.\\
Strader, J., Brodie, J., Forbes, D., 2004, AJ, 127, 295\\
Strader,  J., Brodie, J., Spitler, L., Beasley, M., 2006, AJ, in press\\
Schweizer F., Seitzer P., 1992, AJ 104, 1039 \\
Trager, S., Faber, S., Worthey, G., Gonzalez, J., 2000, AJ, 119, 1645\\ 
Terlevich, A., Forbes, D., 2002, MNRAS, 330, 547 (TF02)\\
Thomas, D., Maraston, C., Bender, R., de Oliveira, C. M., 2005,
ApJ, 621, 673 (T05)\\
Toomre, A., Toomre J., 1972, ApJ, 178, 623\\
Whitmore, B., Schweizer F., 1995, AJ, 109, 960 \\
Whitmore, B., Miller B., Schweizer F., Fall M., 1997, AJ, 114, 797 \\
Whitmore, B., B., Schweizer, F., Kundu, A.,
Miller, B., 2002, AJ, 124, 147 (W02)\\
Zepf, S., Ashman, K., Geisler, D., 1995, ApJ, 443, 570 (Z95)\\

\begin{table*}

\begin{tabular}{lcccccc}
\multicolumn{6}{c}{{TABLE 1} Galaxy central ages}\\
\hline
\hline
Galaxy & TF02 & T05 & D05 & SB06 & Adopted\\
	& (Gyr) & (Gyr) & (Gyr) & (Gyr) & (Gyr) \\ 
\hline
 NGC~584	&2.1	&2.8	&3.8	&5.6 &	3.6\\

 NGC~1052	&	&21.7	&2.9	&	&2.9\\

 NGC~1316	&2.8	&3.2	&	&	&3.0\\

 NGC~1374	&9.8	&	&	&	&9.8\\

 NGC~1379	&7.8	&	&	&	&7.8\\

 NGC~1380	&6.2	&	&	&	&6.2\\

 NGC~1399	&5.0	&	&	&	&5.0\\

 NGC~1404	&5.0	&	&	&	&5.0\\

 NGC~1407	&	&7.4	&2.5	&	&7.4\\

 NGC~1427	&6.5	&	&	&	&6.5\\

 NGC~1700	&2.3	&2.6	&2.6	&5.1	&2.5\\

 NGC~3115	&	&	&2.6	&8.4	&8.4\\
	
 NGC~3377	&4.1	&3.6	&3.6	&5.2	&4.1\\

 NGC~3379	&9.3	&10.0	&10.9	&8.2	&9.6\\

 NGC~3384	&	&	&6.5	&	&6.5\\

 NGC~3610	&	&	&1.8	&	&1.8\\

 NGC~3923	&	&3.3	&2.6	&	&3.0\\

 NGC~4278	&8.4	&12.0	&	&12.4	&10.9\\

 NGC~4365	&	&	&3.6	&7.9	&7.9\\

 NGC~4374	&11.0	&12.8	&3.8	&11.2	&11.7\\

 NGC~4472	&8.5	&9.6	&	&9.6	&9.2\\

 NGC~4494	&	&	&6.7	&	&6.7\\
	
 NGC~4552	&9.6	&12.4	&	&12.3	&11.4\\

 NGC~4594	&	&	&	&10.0	&10.0\\

 NGC~4621	&	&	&	&8.7	&8.7\\

 NGC~4636	&	&	&	&10.0	&10.0\\

 NGC~4649	&11.0	&14.1	&	&	&12.6\\

 NGC~5322	&	&	&2.4	&	&2.4\\

 NGC~5557	&	&	&7.0	&	&7.0\\

 NGC~5846	&12.0	&14.2	&10.2	&8.4	&11.2\\

 NGC~5982	&	&	&12.3	&	&12.3\\

 NGC~6702	&1.9	&1.7	&	&1.6	&1.7\\

 NGC~6868	&15.0	&	&	&	&15.0\\

 NGC~7332	&4.5	&	&1.0	&	&2.8\\

 NGC~7619	&9.0	&15.4	&5.9	&	&10.1\\

 IC~4051		&12.0	&15.0	&	&	&13.5\\

\hline
\end{tabular}

\end{table*}

\begin{table*}

\begin{tabular}{lcccccccccc}
\multicolumn{11}{c}{{TABLE 2} Globular Cluster Mean Colors}\\
\hline
\hline
Galaxy&Type&M$_V$&Blue (V-I)o&error &Red (V-I)o& error &
Ref.&Bimodal&Filter&HST?\\
\hline
 N0584&E4& -21.40&0.96 &0.05 &1.16 &0.05 & K01&L &V-I & Y \\
 N1052&E4 & -20.99&0.90 &0.10 &1.08 &0.10 & F01&Y &B-I & N \\
 N1316&S0p & -22.79&0.90 &0.03 &1.03 &0.03 & G04&Y &V-I & Y \\
 N1374&E1& -20.40&0.94&0.05 &1.15 &0.05 & B06&Y &C-T1 & N \\
 N1379&E0& -20.66&0.95&0.05 &1.13 &0.05 & B06&Y &C-T1 & N \\
 N1380&S0& -21.70&0.87&0.10 &1.16 &0.10 & K97&Y &B-R & N \\
 N1399&cD& -21.85&0.96&0.03 &1.16 &0.03 & F98&Y &B-I & Y \\
 N1404&E1& -21.41&0.91 &0.03 &1.17 &0.03 & F98&Y &B-I & Y \\
 N1407&E0& -20.87&0.93 &0.02 &1.16 &0.02 & F06&Y &B-I & Y \\
 N1427&E5& -20.49& 0.97&0.10 &1.17 &0.10 & F01&Y &C-T1 & N  \\
 N1700&E4& -22.49&0.90 &0.10 &1.12 &0.10 &  B00&Y &B-I & N \\
 N3115&S0& -20.84&0.92 &0.05 &1.15 &0.05 & L01&L &V-I & Y \\
 N3377&E5& -19.79&0.93 &0.05 &1.10 &0.05 & K01&L &V-I & Y \\
 N3379&E1& -20.69&0.96 &0.05 &1.17 &0.05 & L01&L &V-I & Y \\
 N3384&SB0& -20.13&0.94 &0.05 &1.21 &0.05 & L01&L &V-I & Y \\
 N3610&E5& -21.35&0.94 &0.03 &1.16 &0.03 & W02&Y &V-I & Y \\
 N3923& E4& -21.92&1.00 &0.10 &1.20 &0.10 & Z95&Y &C-T1 & N \\
 N4278&E1& -21.11&0.92 &0.05 &1.12 &0.05 & K01&L &V-I & Y \\
 N4365& E3& -21.41&0.98 &0.03 &1.19 &0.03 & L01&L &V-I & Y \\
 N4374& E1& -20.92&0.91 &0.03 &1.07 &0.06 & G03&Y &B-R & N \\
 N4472&E2& -22.81&0.94 &0.03 &1.21 &0.03 & L01&Y &V-I & Y \\
 N4494& E1& -22.00&0.90 &0.03 &1.10 &0.03 & L01&Y &V-I & Y \\
N4552&E5& -21.24&0.95 &0.03 &1.17 &0.03 & L01&Y &V-I & Y \\
 N4594&Sa& -22.00 §&0.94&0.03 &1.18 &0.03 & L01&Y &V-I & Y \\
 N4621&E5 & -21.33&0.95 &0.03 &1.13 &0.03 & K01&Y &V-I & Y \\
 N4636&E & -20.58&0.95 &0.02 &1.19 &0.02 & D05&Y &C-T1 & N \\
 N4649&E2& -22.40&0.95&0.03 &1.21 &0.03 & L01&Y &V-I & Y \\
 N5322&E3& -22.09&0.96 &0.03 &1.14 &0.03 & H06&Y &B-I & Y \\
 N5557&E1& -22.24&0.96 &0.03 &1.18 &0.03 & H06&Y &B-I & Y \\
 N5846&E0& -22.23&0.94 &0.03 &1.15 &0.03 & F97&Y &V-I & Y \\
 N5982& E3& -21.80&0.97 &0.05 &1.16 &0.05 & K01&L &V-I & Y \\
 N6702&E& -21.98&0.83 &0.10 &1.16 &0.10 &  G01&Y &B-I & N \\
 N6868&E& -22.17&0.91 &0.07 &1.12 &0.07 &  D02&Y &B-R & N \\
 N7332&S0p& -20.09&0.88 &0.10 &1.08 &0.10 & F01&L &B-I&  N  \\
 N7619&E2& -22.12&0.99 &0.03 &1.24 &0.03 & L00&Y &V-I & Y \\
 I4051&E& -21.75&0.98 &0.05 &1.16 &0.05 & L00&L &V-I & Y \\

\hline
\end{tabular}

\end{table*}

\begin{figure*}
\includegraphics[angle=0, width=4in]{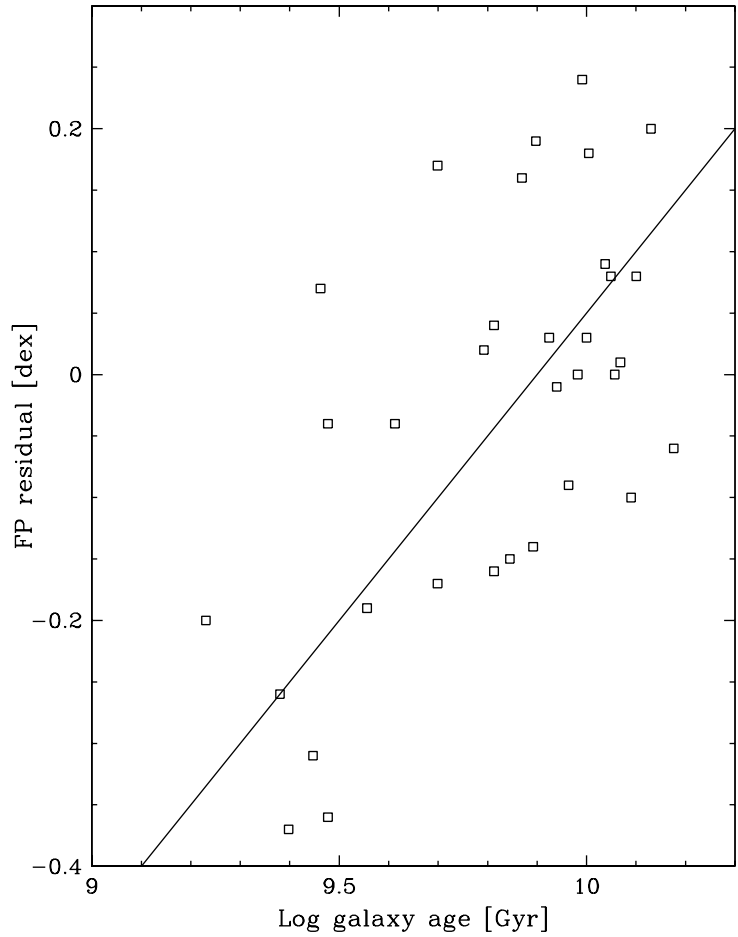}
\caption{\label{fig1} 
Residual from the fundamental plane versus galaxy age. The
deviation from the standard fundamental plane (Prugniel \& Simien
1996) is plotted against our adopted galaxy age. The solid line
is the fit from Forbes, Ponman \& Brown (1998) for a larger
sample of elliptical galaxies. The sample used here follows the
general Forbes \etal trend with a rms scatter in log age of 0.2
dex (or 1.6 Gyrs). 
}
\end{figure*}

\begin{figure*}
\includegraphics[angle=0, width=4in]{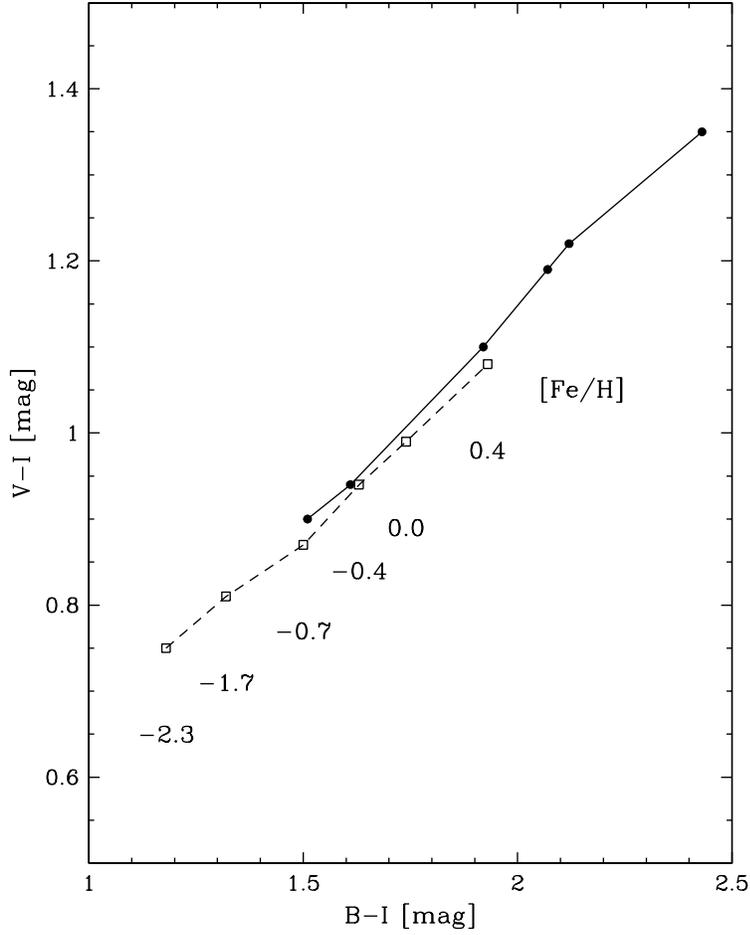}
\caption{\label{fig2} 
V--I vs B--I single stellar population models. From the SSP
models of BC03 we show the single age tracks for a 12.5 Gyr old
population (solid line) and a 2 Gyr popultion (dashed line). Six
different metallicities from [Fe/H] = --2.3 to +0.4 dex are
indicated. For typical B--I colors
of GCs (i.e. 1.5 $<$ B--I $<$ 2.2) the predicted V--I
color is $\le$ 0.03 mag. redder if GCs are as young as 2 Gyrs
compared to an assumption of 12.5 Gyrs.
}
\end{figure*}

\begin{figure*}
\includegraphics[angle=-0, width=4in]{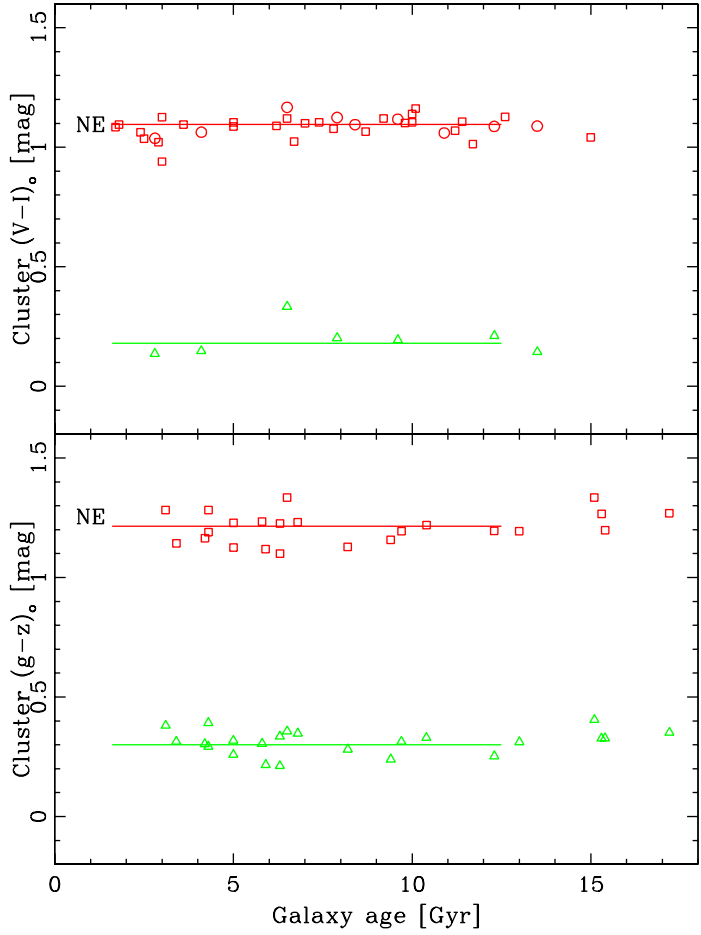}
\caption{\label{fig3} Mean GC colors vs.~central galaxy 
age for early-type galaxies. The red open squares (bimodal = Yes)
and circles (bimodal = Likely) show the mean colors of the 
red GC subpopulation and the green trainagles show the  
color difference between the red and blue subpopulations.  
The GC colors have been corrected for the
GC color--galaxy luminosity relation.  The 
solid lines show the predicted color for a 
12.5 Gyr old [Fe/H] = --0.7 single stellar population; called the no
evolution (NE) track. The data are consistent
with having a constant color and hence the no evolution (NE)
track.
}
\end{figure*}

\begin{figure*}
\includegraphics[angle=-0, width=4in]{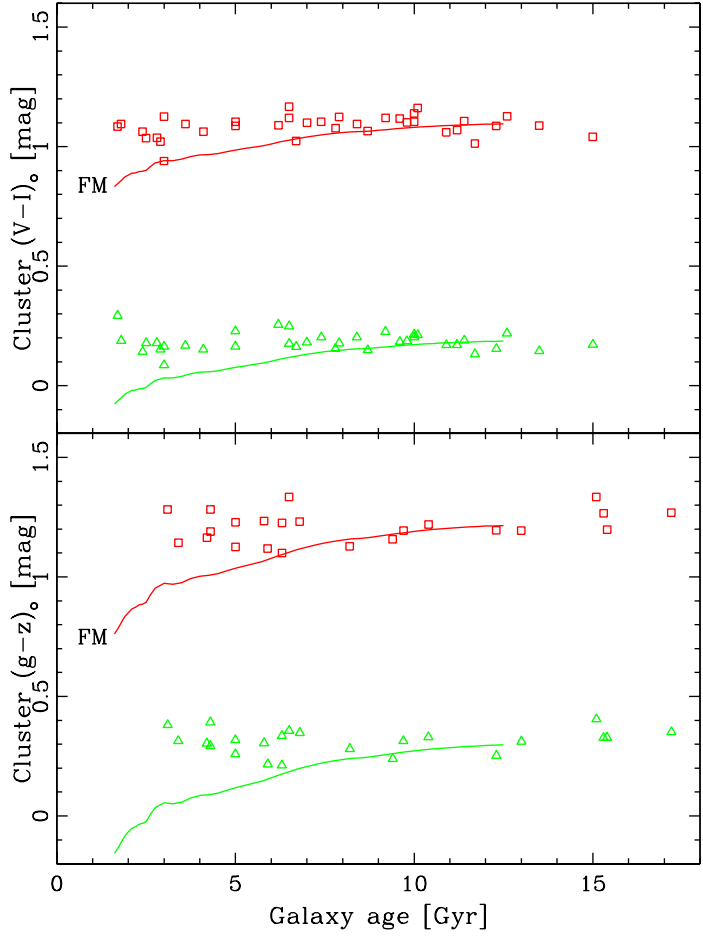}
\caption{\label{fig4} 
Mean GC colors vs.~central galaxy 
age for early-type galaxies. 
Same symbols as in Figure 2. The 
solid lines show the 
predicted color evolution for a fixed 
metallicity of [Fe/H] = --0.7; called the fixed metallicity (FM) 
track.  The data are inconsistent
with the fixed metallicity (FM) track.
}
\end{figure*}

\begin{figure*}
\includegraphics[angle=-0, width=4in]{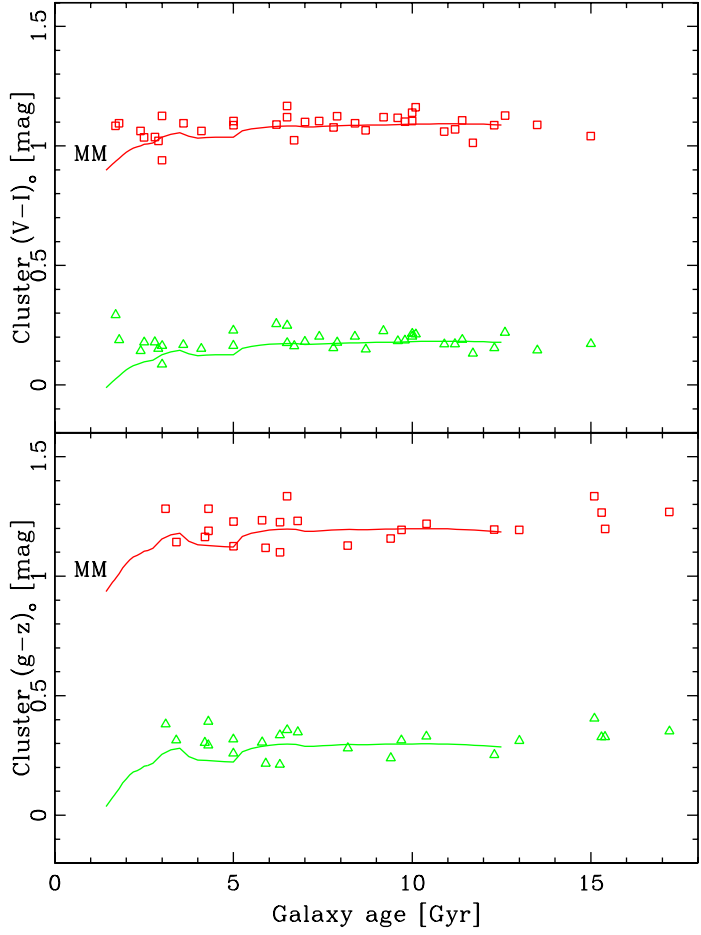}
\caption{\label{fig5} 
Mean GC colors vs.~central galaxy 
age for early-type galaxies. 
Same symbols as in Figure 2. The 
solid lines show the 
predicted color evolution for an 
age-metallicity relation based on the Galactic disk (see text for details);  
called the mixed metallicity (MM) 
track.  The data are consistent
with the mixed metallicity (MM) track.
}
\end{figure*}

\end{document}